\documentclass{aa}  

\usepackage{graphicx}
\usepackage{txfonts}
\usepackage[colorlinks=true]{hyperref}
\hypersetup{linkcolor=blue, citecolor=blue}
\usepackage[draft]{todonotes}
\usepackage{mathtools}
\usepackage{booktabs}
\usepackage{multirow}
\usepackage{bbold}
\usepackage{caption}
\usepackage{subcaption}
\usepackage{bm}

\usepackage{xcolor}
\definecolor{vaso}{rgb}{1.0, 0.0, 0.0}
\definecolor{alex}{rgb}{0.0, 0.0, 1.0}

\newcommand{\B}{\mathbf{B}}
\newcommand{\bfphi}{\bm{\varphi}}
\newcommand{\kbf}{\mathbf{k}}
\newcommand{\xbf}{\mathbf{x}}
\newcommand{\Bx}{\mathbf{B}(\xbf)}

\begin{document}

   \title{Reconstructing Galactic magnetic fields from local measurements for backtracking ultra-high-energy cosmic rays}

    \titlerunning{Reconstructing GMFs from local data for backtracking UHECRs}

   \author{Alexandros Tsouros \inst{1} \fnmsep \inst{2}\thanks{tsouros@physics.uoc.gr}, Gordian Edenhofer  \inst{3}\fnmsep\inst{4}, Torsten Enßlin \inst{3}\fnmsep\inst{4}, Michalis Mastorakis \inst{1} \fnmsep \inst{2} , Vasiliki Pavlidou \inst{1} \fnmsep \inst{2}
          }

   \institute{
        Department of Physics \& ITCP, University of Crete, GR-70013, Heraklion, Greece
        \and
        Institute of Astrophysics, Foundation for Research and Technology-Hellas, Vasilika Vouton, GR-70013 Heraklion, Greece
        \and Ludwig Maximilian University of Munich, Geschwister-Scholl-Platz 1, 80539 Munich, Germany
        \and Max Planck Institute for Astrophysics, Karl-Schwarzschild-Stra{\ss}e 1, 85748 Garching, Germany
        }
    \authorrunning{A. Tsouros et al. 2023}
    
   \date{Received ; accepted }

 
  \abstract
   {Ultra-high energy cosmic rays (UHECRs) are highly energetic charged particles with energies exceeding $10^{18}$ eV. These energies are far greater than those achieved in Earth-bound accelerators, and identifying their sources and production mechanism can shed light into many open questions in both astrophysics as well as high energy physics. However, due to the presence of the Galactic magnetic field (GMF) they are deflected and hence the location of their true source on the plane of the sky (PoS) is concealed. The identification of UHECR sources is an open question, excacerbated by the large uncertainties in our current understanding of $3$-dimensional structure of the GMF. This difficulty arises from the fact that currently all GMF observations are integrated along the line-of-sight (LoS). However, thanks to upcoming stellar optopolarimetric surveys as well as Gaia data on stellar parallaxes, we expect that local measurements of the GMF in the near future will become available.}
   {The question then arises: given such a set of (sparse) local GMF measurements, what is the optimal way to use them in backtracking UHECRs through the Galaxy? In this paper, we evaluate the reconstruction of the GMF, in a limited region of the Galaxy, through Bayesian inference using principles of Information Field Theory.}
   {We employ methods of Bayesian statistical inference in order to estimate the posterior distribution of the GMF configuration within a certain region of the Galaxy from a set of sparse simulated local measurements. Given the energy, charge, and arrival direction of a UHECR, we can backtrack it through GMF configurations drawn from the posterior, and hence calculate the probability distribution of the true arrival directions on the PoS, by solving the equations of motion for each case. 
   }
   {We show that, for a weakly turbulent GMF, it is possible to correct for its effect on the observed arrival direction of UHECRs to within $\sim 3^\circ$. For completely turbulent fields, we show that our procedure can still be used to significantly improve our knowledge on the true arrival direction of UHECRs.}
   {}


   \maketitle
%
\section{Introduction}
The identification of ultra high energy cosmic ray (UHECR) sources remains one of the central questions in modern high-energy astrophysics. A possible resolution to this problem would shed light on the astrophysical mechanism that produces UHECRs, as well as their composition (either protons or heavier nuclei), which at high energies still remains a subject of debate. Knowledge of their composition may, in turn, provide insights into high-energy particle phenomenology at energies not yet probed by particle accelerators (\citealt{pavlidou_tomaras}; \citealt{rom&tom&pav1}; \citealt{rom&tom&pav2}).

Despite the fact that several theoretical candidates for UHECR sources have been proposed (\citealt{bhattacharjee}; \citealt{TorresReview}), the identity of their sources remains elusive. The primary reason for this is that UHECRs are charged particles and are hence deflected by the Galactic magnetic field (GMF) as well as the intergalactic magnetic field. Even if several of the detected UHECR originate from a single nearby cosmic-ray-bright source \citealt{Auger-TA}, their arrival directions would be very spread out on the sky, and any residual clustering would be centered off-source, due to magnetic deflections. This is unlike the case of photons or neutrinos, where positional identifications events with their likely sources can be attempted even with very low-number statistics. 

The main difficulty in resolving the GMF stems from the fact that $3D$ tomographic realizations of the intervening magnetic fields are notoriously difficult to acquire. Specifically for the GMF, most observables that are currently available are integrated along the line of sight (LoS). Due to this limitation, and specifically for the GMF, the principal approach (\citealt{Takami}) is relying on GMF models that are acquired by parameter fitting three separate components - a toroidal, a poloidal, and a Gaussian random field (\citealt{JF12}; \citealt{JF12-2}; \citealt{Sun10}; \citealt{Sun10-2}).

Nonetheless, it is possible to acquire direct information regarding the $3D$ structure of the GMF. Gaia data on stellar distances have localised more than a billion stars in the Galaxy by accurately determining stellar parallaxes (\citealt{Gaia1}; \citealt{Gaia-2}; \citealt{Bailer-Jones}). This data - in addition to other available spectroscopic data - has been used in order to construct $3D$ tomographic maps of the dust density distribution of certain Galactic regions (\citealt{Green}; \citealt{Lallement}; \citealt{LeikeEnsslin}; \citealt{LeikeEnsslin2}). These reconstructions, however, do not constrain the magnetic field which is of primary interest in cosmic ray physics. 

Nonetheless, there do exist probes that provide information regarding the structure of the GMF in $3D$. For instance, the linear polarization of starlight is of particular interest; while starlight starts off unpolarized from its source, it will usually acquire a linear polarization until it is observed on Earth, due to dichroic absorption by dust particles aligned with the ambient magnetic field (\citealt{Andersson}). 

Upcoming optopolarimetric surveys, such as PASIPHAE (\citealt{Pasiphae}; \citealt{WALOP-south1}; \citealt{WALOP-south2}; \citealt{Southpol}), are expected to provide a large number of high quality stellar polarization measurements on more than a million of stars. With stellar distances also known from the Gaia survey, this data can be used to provide direct tomographic measurements of the plane of the sky (PoS) component of the GMF at the location of the dust clouds (\citealt{Panopoulou}; \citealt{Pelgrims}; \citealt{Davis}; \citealt{Chandra}; \citealt{ST}). Used jointly with available LoS information (see for example \citealt{Tahani2022a}, \citealt{Tahani2022b}), we can expect to have local and sparse data of the GMF in the near future, that could be used in order to provide a $3D$ tomographic map of a particular region of interest. Given such a map, one can then backtrack UHECRs through that region, thus improving the localisation of their source on the sky, modulo the effects of the intergalactic magnetic field. To be more specific, one might be interested in reconstructing the region of the GMF through which the UHECRs `hotspots' (\citealt{Kawata};\citealt{Abbasi};\citealt{AugerHotspot}) have to travel through.


In this paper, we will address the problem of reconstructing the posterior density function (PDF) for the true arrival directions of UHECRs, using local, sparse, and (statistically) uniformly distributed GMF measurements. In essence, we are interested in solving the inverse problem in a Bayesian setting, wherein one is given local data and is tasked with calculating the posterior distribution for the GMF configurations in the region of interest. For Bayesian field inference, information field theory (IFT) was developed \citealt{IFT0}, and demonstrated in a number of contexts \citealt{IFT-AI}. Here, we will use an adapted version of the algorithm used in \citealt{Leike}, after we generalise it in order to make it applicable to divergence-free vector fields. Using the reconstructed posterior distribution over GMF realizations, we correct for the effect of the GMF on the observed arrival direction of the UHECR. 

The work is structured as follows: in section \ref{sec:methods} we describe the Bayesian setting that will be employed, and motivate the principal components of the algorithm that forms the core of our reconstruction scheme, as well as describe our assumptions. In section \ref{sec:results} we present the main results of this paper, by applying our reconstruction scheme on magnetic fields that are locally and sparsely sampled, for different relative strengths of the turbulent and the uniform component, as well as different sampling intervals. In section \ref{sec:conclusion} we summarize and discuss our conclusions. 

\section{Methodology} \label{sec:methods}

In Bayesian inference for continuous signals, we are in general interested in reconstructing the posterior probability distribution $\varphi(\xbf)$ defined over a domain $\mathcal{V}$ from a given data set $d$: 
\begin{equation}\label{Bayes}
    P(\varphi|d) \propto P(d|\varphi) P(\varphi)
\end{equation}
The distribution $P(d|\varphi)$ is the \emph{likelihood}, and quantifies how likely acquiring $d$ as our data is, given the configuration is $\varphi(\xbf)$. The function $P(\varphi)$ is the \emph{prior probability}, containing our information on the $\varphi(\xbf)$ before $d$ is taken into account. The proportionality factor in \eqref{Bayes} is determined by normalising the posterior to unity.

Mathematically, $\varphi(\xbf)$ is a continuous function, and the distributions $P(\varphi)$, $P(\varphi|d)$, and $P(d|\varphi)$ are functionals. For example, an important case is the Gaussian distribution, 
\begin{equation}\label{Gaussian_sc}
     \mathcal{G}(\varphi-m,\varPhi) \equiv \frac{\exp \left[ -\frac{1}{2}(\varphi-m) \varPhi^{-1}(\varphi-m) \right]}{|2 \pi \varPhi|^{\frac{1}{2}}},
\end{equation}
where $m(\xbf) = \langle \varphi(\xbf) \rangle$ is the mean. The quantity $\varPhi(x_1,x_2) = \langle \delta\varphi(x_1) \delta\varphi(x_2) \rangle$ is the covariance of the distribution, where we have defined  $\delta\varphi(x) \equiv \varphi(x) - m(x)$. $|\varPhi|$ denotes the determinant of $\varPhi$. In equation \eqref{Gaussian_sc}, note that there is also an implicit integration, that is, 

\begin{align*}
    (\varphi-m) \varPhi^{-1}&(\varphi-m) \equiv  \\ &\int  d^3 x d^3 x' [\varphi(\xbf)-m(\xbf)] \varPhi^{-1}(\xbf,\xbf')
     [\varphi(\xbf')-m(\xbf')].
\end{align*}
In order to avoid cluttered notation, this integration will always be written implicitly.


\subsection{Prior} \label{sec:prior}


The signal that we are interested in reconstructing is the GMF, $\mathbf{B}(\xbf)$, which is a vector quantity. Following equation \eqref{Gaussian_sc}, the prior is written as 

\begin{equation}\label{Gaussian_vec}
    \mathcal{G}(\delta \mathbf{B}(\xbf) , M) \equiv \frac{\exp \left[ -\frac{1}{2}\delta B_i M_{ij}^{-1}\delta B_j \right]}{|2 \pi M|^{\frac{1}{2}}},
\end{equation}
where $\delta B_i(\xbf) = B_i(\xbf) - \langle B_i(\xbf) \rangle$, with $\langle B_i(\xbf) \rangle$ denoting the mean of the distribution.

Here the indices correspond to individual components of the vector $\mathbf{B}(\xbf)$, and the Einstein summation convention is assumed, and $\mathbf{B_0}$ is the mean field. Here, the covariance matrix inherits two indices since by definition

\begin{equation} \label{GMFcovariance}
    M_{ij} (\xbf, \xbf') \equiv \langle \delta B_i (\xbf) \delta B^*_j (\xbf') \rangle,
\end{equation}

In lack of information regarding the geometry and statistics of the GMF in the region of interest, we model the prior as being statistically isotropic and homogeneous. In Fourier space, this is equivalent to assuming that the two-point correlation function \eqref{GMFcovariance} takes the form 

\begin{equation} \label{2pointcor}
    \langle \delta \hat{B_i} (\mathbf{k}) \delta \hat{B}^*_j (\mathbf{k}') \rangle= \frac{(2 \pi)^3}{2} \mathcal{P}_{ij}(\mathbf{k})\delta^{(3)}(\mathbf{k}-\mathbf{k}')P(k),
\end{equation}
where 

\begin{equation} \label{projector}
   \mathcal{P}_{ij}(\mathbf{k}) \equiv \delta_{ij} - \hat{k}_i \hat{k}_j,
\end{equation}
is the transverse projection operator, $\hat{k}_i$ denotes the unit $\mathbf{k}$-vector in the $i$-th direction, $\delta_{ij}$ is the kronecker delta, $\delta^{(3)}(\mathbf{k}-\mathbf{k}')$ is the three dimensional Dirac delta function, and $\delta \hat{B}_j^*(\mathbf{k})$ denotes the complex conjugate of $\delta \hat{B}_j(\mathbf{k})$. Further, the norm of the $\mathbf{k}$-vector is henceforth denoted as $k$. The function $P(k)$ is the magnetic power spectrum, and needs to be inferred as well. Equation \eqref{2pointcor} assumes that no prior expectation on   the presence of helicity of any sign exists\footnote{Helicity would add a complex, antisymmetric term of the form $i \varepsilon_{ijl} k_l H(k)$, with $H(k)$ the real helicity spectrum, which satisfies
$-P(k) < H(k) < P(k)$.}. We note that magnetic helicity in the reconstructed magnetic field is not excluded thereby, it just needs to be requested by the data and will not be enforced by the prior.

Instead of inferring $\mathbf{B}(\xbf)$ directly, it is more practical to work with a latent vector field $\bfphi (\xbf)$ that does \emph{not} satisfy the divergence free condition, and thus contains more degrees of freedom than needed. The Fourier space correlation structure for $\bfphi$ will be assumed to be isotropic and homogeneous, and this will take the form  

\begin{equation} \label{iso-hom}
    \langle \delta \hat{\varphi}_i(\mathbf{k}) \delta \hat{\varphi}^*_j(\mathbf{k}') \rangle = (2 \pi)^3 \delta_{ij}\delta^{(3)}(\mathbf{k}-\mathbf{k}')P(k),
\end{equation}
with the understanding that $\mathbf{B}(\xbf)$ and $\bfphi (\xbf)$ related to each other by an application of $\mathcal{P}_{ij}(\mathbf{k})$ by

\begin{equation} \label{def_bf}
   \hat{B}_i(\kbf) \equiv \frac{3}{2} \mathcal{P}_{ij}(\kbf)\hat{\varphi}_j (\kbf) = \frac{3}{2} \left( \hat{\varphi}_i(\mathbf{k}) - [\hat{\varphi}_j(\kbf) k_j] \frac{k_i}{k^2} \right).
\end{equation}
in Fourier space. This definition is such that the resulting field $\B$ is guaranteed to be divergence free, since then $k_i \hat{B}_i(\kbf) = 0$ in harmonic space, implying $\nabla \cdot \B = 0$ as required. The $3/2$ factor compensates for the loss of power that the subtraction of degrees of freedom causes. This is justified by the original a priori assumption of statistical isotropy for $\bfphi$, which leads to equipartition along the three directions in Fourier space. Equation \eqref{def_bf} is in accordance with the correlation structure assumed in equation \eqref{2pointcor} (\citealt{Hammurabi}). 

The power spectrum $P(k)$ is unknown and needs to be inferred as well. It is modeled as the sum of a power law component and an integrated Wiener process component (for details regarding the generative model for power spectra, the reader should refer to \citealt{variable_shadow}). The parameters that define it, as well as their respective prior PDFs are 

\begin{enumerate}
    \item The total field offset for all field components. This controls the mean value around which the random field fluctuates, $\mathbf{B_0}$. We set a value of $0$, which means that in the absence of data, the mean overall field is assumed to be zero-centered. 

    \item The standard deviation of the total offset. This is a random variable with a prior log-normal distribution. We set its mean at $3$ $\mu$G and its own standard deviation at $1$ $\mu$G. This reflects our current understanding of typical GMF values in the interstellar medium that a (randomly oriented) mean magnetic field of this typical strength within the reconstructed volume is conceivable, but not enforced.

    \item The total spectral energy\footnote{Here, the term `energy' is used in the context of signal processing, and it does not refer to physical energy, although the two are related.}. This controls the amplitude of the fluctuations in configuration space. Its prior probability distribution is log-normal with a mean and standard deviation set to unity. This is essentially the standard deviation of the field values calculated over the complete set of voxels.
    
    \item The spectral index; the exponent of the pure power law component. Its prior distribution is normal, with a mean set as the Kolmogorov index, $-11/3$, and we assume unit standard deviation. 
    
    \item Amplitude of the integrated Wiener process component, controlling the deviations from a pure power-law.

    Its prior distribution is log-normal, with mean and standard deviation equal to $1.5$ and $1$ respectively. These parameters are chosen so as to allow for deviations from a pure power law without destroying the approximately power-law overall behaviour of the power spectrum.

\end{enumerate}
The parameters of the amplitude model are assumed to be relative to a unit-less power spectrum, i.e. the parameters are assumed to be agnostic to changes in the volume of the target subdomain $\mathcal{V}^3$. In table \ref{priorparams}, we summarise our choice for the probability distributions for the parameters of the power spectrum,  as well as their respective means and standard deviations. 

\begin{table*}
  \caption[]{Parameters that define the power spectrum prior}
     \label{priorparams}
     \centering
     \begin{tabular}{c c c c}
        \hline\hline
        Parameter & Distribution & Mean & Standard deviation \\
        \hline
        Total offset ($\mathbf{B_0}$)& Not-applicable & $0$ & Not-applicable  \\
        Total offset st. dev. & Log-normal  & $3$ $\mu$G & $1$ $\mu$G \\
        Total spectral energy  & Log-normal & $1$ $\mu$G & $1$  $\mu$G  \\
        Spectral index  & Normal & $-\frac{11}{3}$ & $1$ \\
        Int. Wiener process amplitude & Log-normal & $1.5$ & $1$ \\ 
        \hline
     \end{tabular}
\end{table*}

In closing the discussion on the prior, we consider the implementation of anisotropy. In general, the Galactic magnetic field in the region within which we wish to reconstruct it, will be assumed to admit the two-component structure 

\begin{equation} \label{uni+fluct}
    \Bx = \mathbf{B}_0 + \mathbf{b}(\xbf),
\end{equation}
where $\mathbf{B}_0$ is a uniform field parallel to the PoS. Since the observed arrival velocity of the UHECR will be chosen to be parallel to the LoS (see section \ref{sec:results}), the PoS component of $\mathbf{B}_0$ will dominate the UHECR's deflection and the LoS component will be hardly constrained. Additionally, $\mathbf{b}(\xbf)$ is a fluctuating field with zero mean, which can be physically interpreted as the turbulent field. \emph{Our prior is agnostic to the direction and magnitude of $\mathbf{B}_0$, and it is a central task to infer it from the data}. The relative strength of the two components will be quantified by the turbulent-to-uniform ratio $\lambda$, defined as 

\begin{equation}
    \lambda \equiv \frac{b_{\text{rms}}}{B_0},
    \label{lambda}
\end{equation}
where $B_0$ is the norm of $\mathbf{B}_0$, and $b_{\text{rms}}$ is the root mean square value of the fluctuating component's magnitude, in the domain $\mathcal{V}$. For $\lambda \gg 1$ we have \emph{strong turbulence}, for $\lambda \ll 1$ we have \emph{weak turbulence}, and for $\lambda \simeq 1$ we have \emph{intermediate turbulence}. As we will demonstrate, $\lambda$ is the main parameter that controls how well the sources of UHECRs can be localized by the use of local GMF data along, of course, with the sampling rate.

\subsection{Likelihood} \label{sec:likelihood}

In order to construct the likelihood, we consider how the data is acquired from the true signal $\bm{B}$. We will assume that the $i$-th datapoint is given by a measurement process of the form

\begin{equation}
    \mathbf{d}^{(i)} = \int R(\xbf,\xbf_i) \mathbf{B}(\xbf) d^3x + \mathbf{n}^{(i)}.
\end{equation}
The operator $R$ is a map from the signal space to the data space. For this proof of concept work, we will make the simplifying assumption that local measurements of field components are possible, {\it i.e.} $R(\xbf,\xbf_i)= \delta^{(3)}(\xbf - \xbf_i)$. In practice, most astrophysical measurements of magnetic fields (Faraday rotation, UHECR deflection, etc.)  provide at least a LoS, particle trajectory, or even sub-volume averaged field strength. This only complicates the numerical reconstruction of fields, but does not imply conceptual changes in the formalism.

Thus, since we are assuming local measurements, $R$ is here assumed to be a mask operator.  In practice, since the mock signal is discretised into voxels, $R$ is simply a very sparse matrix. The full vector $\bm{d}$ is therefore a concantenation of a number of $3D$ vectors known on a finite number of points inside the domain wherever we have measurements, and undefined anywhere else. Simply put, then, the operator $R$ merely acts a selection operator that picks out the values of $\mathbf{B}(\xbf)$ where we have measurements. Physically, the assumption is that at the positions where GMF measurements can be obtained, all three components of $\mathbf{B}(\xbf)$ can be measured. This can be accomplished, e.g., by Zeeman observations for the LOS component of the magnetic field. The POS component of the magnetic field can be recovered through a combination of stellar optopolarimetry and stellar distance measures. Optopolarimetric measurements of stars of known distances can be decomposed to the contributions of individual clouds along the line of sight \citealt{Pelgrims}. Then, the average direction of polarization angles induced on starlight by dichroic absorption due to a single cloud reveals the direction of the magnetic field in that cloud. The dispersion of these directions can yield a measurement (within a factor of 2) of the magnetic field strength in the same cloud (\citealt{Chandra}; \citealt{Davis}; \citealt{ST}; \citealt{ST2}).


Additionally, the vector $\mathbf{n}^{(i)}$ is random noise added to the $i$-th \emph{noiseless} data point. It is assumed to be drawn from a Gaussian distribution with standard deviation equal to half the root-mean-square magnitude of the mock signal field. Physically, $\mathbf{n}^{(i)}$ represents the measurement error, assuming an average signal-to-noise ratio (SNR) of $2$.

The likelihood is then calculated by marginalizing over the noise:

\begin{align}
    P(\bm{d}|\bm{B}) &= \int \mathcal{D}\bm{n} P(\bm{d},\bm{n} | \bm{B}) = \int \mathcal{D}\bm{n}P(\bm{d}|\bm{n},\bm{B})P(\bm{n}|\bm{B}) \nonumber  \\
    &= \int \mathcal{D}\bm{n} \delta(\bm{d} - (R \bm{B}+\bm{n}))P(\bm{n}|\bm{B}) \nonumber \\
    &= \mathcal{G}(\bm{d} - R\bm{B},N).
\end{align}
where $N$ is the noise covariance, and we used the shorthand notation $R \bm{B} \equiv  \int R(\xbf,\xbf_i) \mathbf{B}(\xbf) d^3x$. We can write the likelihood in terms of $\bfphi$ by absorbing $\mathcal{P}_{ij}$ into $R$, that is, $R' \equiv R \mathcal{P}_{ij}$. Then, the likelihood becomes

\begin{equation}
    P(\bm{d}|\bfphi) = \mathcal{G}(\bm{d} - R'\bm{\varphi},N).
\end{equation}

\subsection{Approaching the posterior: Geometric variational inference} \label{sec:geoVI}

In the previous two sections, we described how we model the prior and the likelihood for our inference setting. From equation \eqref{Bayes}, the posterior is proportional to the product of the two.

Due to the fact that the magnetic power spectrum $P(k)$ needs to be inferred along with the configuration of the GMF, this inference problem is nonlinear. A way to see this, is that the magnetic field $\mathbf{B}(\xbf)$ couples to $P(k)$ through equations \eqref{Gaussian_vec} and \eqref{2pointcor}. Moreover, the prior PDFs for the parameters that determine $P(k)$ are not Gaussian (with the exception of the spectral index), which is an additional source of nonlinearity. Further, there is no small parameter that could be used in a perturbative analysis about a linear inference case. For this reason, a non-perturbative scheme, called geometrical variational inference (geoVI) developed by \citealt{geoVI} will be utilised. In this section we motivate the basic premises of geoVI. 

The idea is to approximate the true posterior, $P$, with an approximate one, $Q$. The approximate posterior $Q$ is chosen such that the Kullback-Leibler divergence (\citealt{KLdiv}) 
\begin{equation} \label{KL}
    D_{KL}(Q,P) \equiv \int dQ \log \left( \frac{Q}{P}\right)
\end{equation}
between the actual posterior $P$ and an approximate posterior $Q$ is minimized. The main idea of geoVI is to achieve this minimization in a new coordinate system, chosen such that $P$ - in the new coordinate system - locally closely resembles a normalized standard distribution. Once this is done, the approximating posterior $Q$ is chosen to be of the form \eqref{Gaussian_vec}. Then, the mean and covariance are chosen as the parameters with respect to which the KL divergence is minimized.

A few more details on geoVI can be found in Appendix A and in \citealt{geoVI}. Essentially, the algorithm provides approximate posterior samples that can follow the non-Gaussian structure of the posterior to a certain degree. geoVI can be invoked by the Numerical Information Field Theory (NIFTy) package in Python (\citealt{nifty1}; \citealt{nifty3}; \citealt{asclnifty5}). The input that is required is the likelihood and the prior of the original physical model, as described in sections \ref{sec:likelihood} and \ref{sec:prior} respectively.



\section{Results}\label{sec:results}

\subsection{General Procedure} \label{sec:gp}

We are now ready to construct a number of representative examples, in which we will apply the procedure described in the previous section, in order to reconstruct various assumed GMF geometries from a set of local, sparse, noisy, $3D$ GMF mock observations. All of the examples that we will study are created according to the following scheme:

\begin{enumerate}
     \item Define the domain: We choose a cube $\mathcal{V}$ of side length $L = 3$ kpc with periodic boundary conditions (topologically a $3$-torus), and evenly divide it into $N^3$ voxels. The number $N$ defines our resolution. 

    \item Produce a power spectrum $P(k)$ by randomly sampling each of its defining parameters from its respective distribution (see section \ref{sec:prior}). 
    \item Produce the latent field $\bm{\varphi}$ with a correlation structure dictated by equation \eqref{iso-hom}, with it taking a constant value over each voxel.
    
    \item Create the fluctuating part of the synthetic signal in Fourier space, by acting on $\bm{\varphi}$ with the transverse projection operator;
    
    \begin{equation*}
        b_i(\kbf) =  \mathcal{P}_{ij}(\kbf)\varphi_j(\kbf).
    \end{equation*}
    After transforming into configuration space and computing the RMS value of the field, $b_\text{rms}$, we add the uniform component that lies along the $x-y$ plane (serving as our PoS) of a magnitude $\lambda^{-1} b_{\text{rms}}$, for a desired value of $\lambda$. Rescale the total field such that the total RMS magnitude is $5$ $\mu$G.
    
    \item Act on the mock signal with a mask operator, which uniformly chooses a specified small fraction of voxels from the subdivided domain. This is done as follows: if $N_d$ is the (approximate) number of data points we wish to have, then we act on the original array with an operator which masks each voxel with a probability of $1-N_d/N$. Roughly, then, the number of voxels that survive is $N_d$, by construction.

   However, it is more useful to refer to the \emph{mean} sampling `rate'; since by construction the distribution of data points is spatially homogeneous, the mean distance between the data points, as a function of $N_d$, is 
    
    \begin{equation*}
        \ell = \frac{L}{N_d^{1/3}}.
    \end{equation*}
    The mean sampling rate, is thus defined as
    
    \begin{equation}
        k_{\text{sample}} \equiv \ell^{-1} = \frac{N_d^{1/3}}{L}.
        \label{ksample}
    \end{equation}
    
    In physical reality, the data points will be located wherever HI clouds exist in the region under study, which are not positioned uniformly within $\mathcal{V}$, and so the sampling rate will vary with vertical distance from the Galactic plane. Furthermore, there are also non-local measurements which, for example, average over a LoS. In future work, we shall also consider such effects. 
    
    \item Add noise: to each data vector created in the previous step, we add a random vector sampled from a multivariate Gaussian distribution meant to represent observational error. The variance of the distribution is
    
    \begin{equation} \label{noise_cov}
        \sigma^2 = \frac{B^2_{\text{rms}}}{4} = \frac{25}{4} (\mu \text{G})^2
    \end{equation}
    where $B_{\text{rms}}$ is the RMS value of the total magnetic field's norm within $\mathcal{V}$, which is set to $5$ $\mu$G (see step 4). The denominator on the right hand side is chosen such that the average signal to noise ratio is $2$. 

    \item Apply the geoVI algorithm on the data set (see section \ref{sec:geoVI}). The output of the algorithm are samples of the posterior distribution.
    
\end{enumerate}

\begin{figure}
        \centering
        \begin{subfigure}{.5\textwidth}
         \centering
         \includegraphics[scale = 0.18]{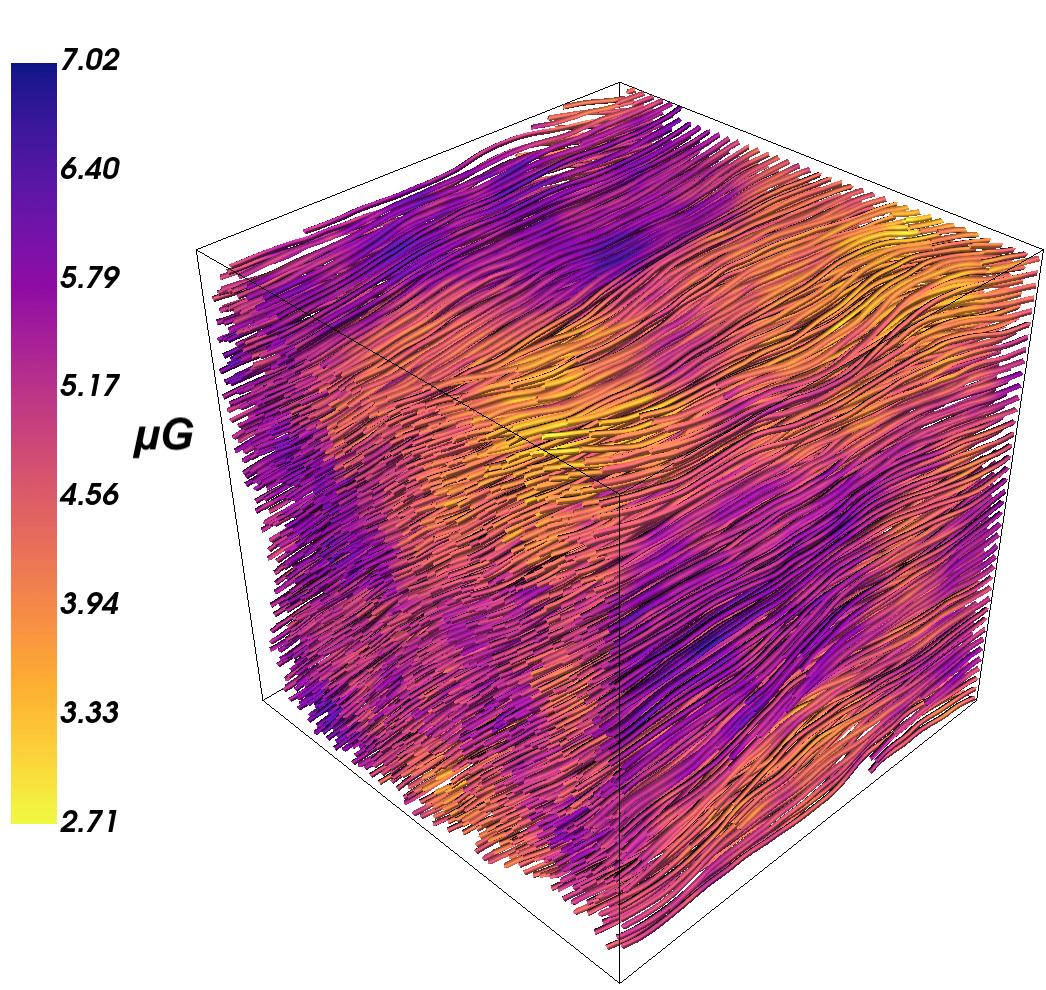}
         \caption{}
         \label{fig:10a}
     \end{subfigure}
     \hfill
     \begin{subfigure}{.5\textwidth}
         \centering
         \includegraphics[scale = 0.18]{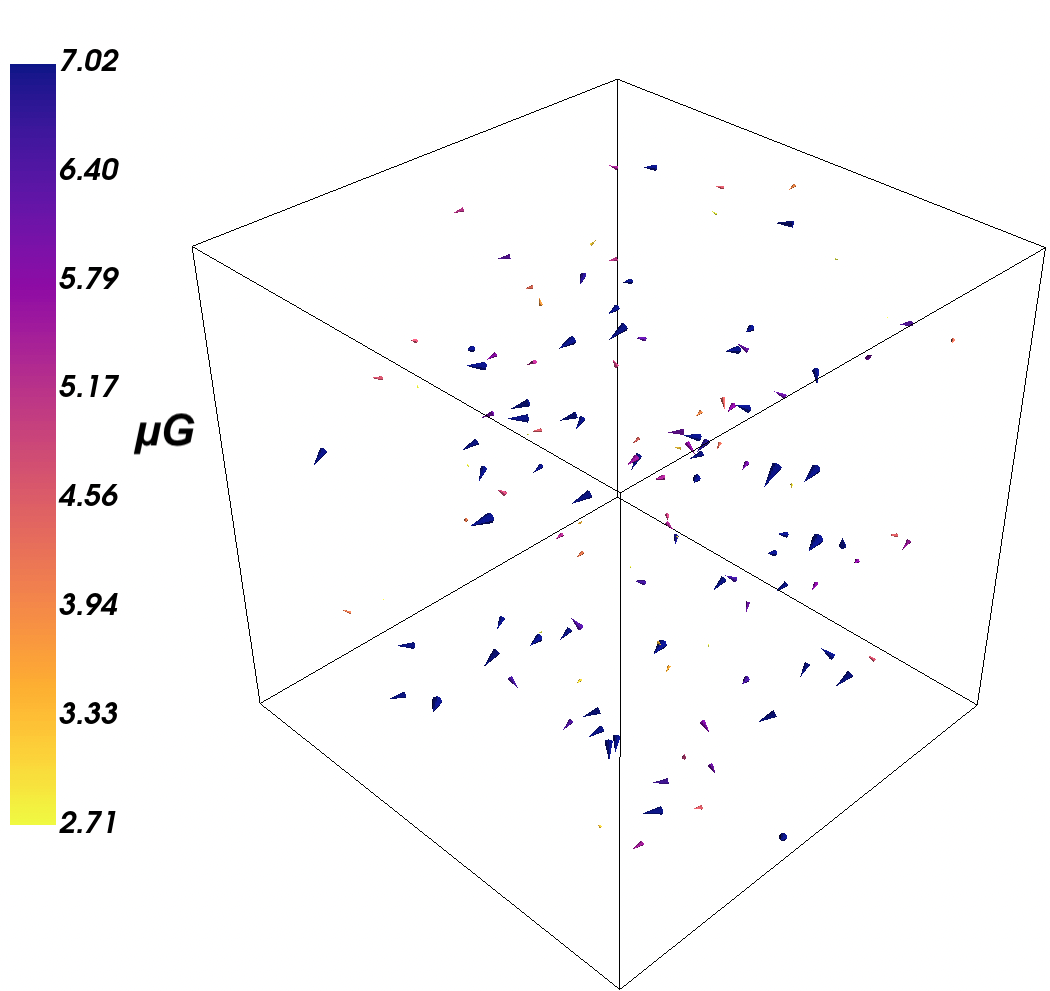}
         \caption{}
         \label{fig:10b}
     \end{subfigure}
     \hfill
     \begin{subfigure}{.5\textwidth}
         \centering
         \includegraphics[scale = 0.18]{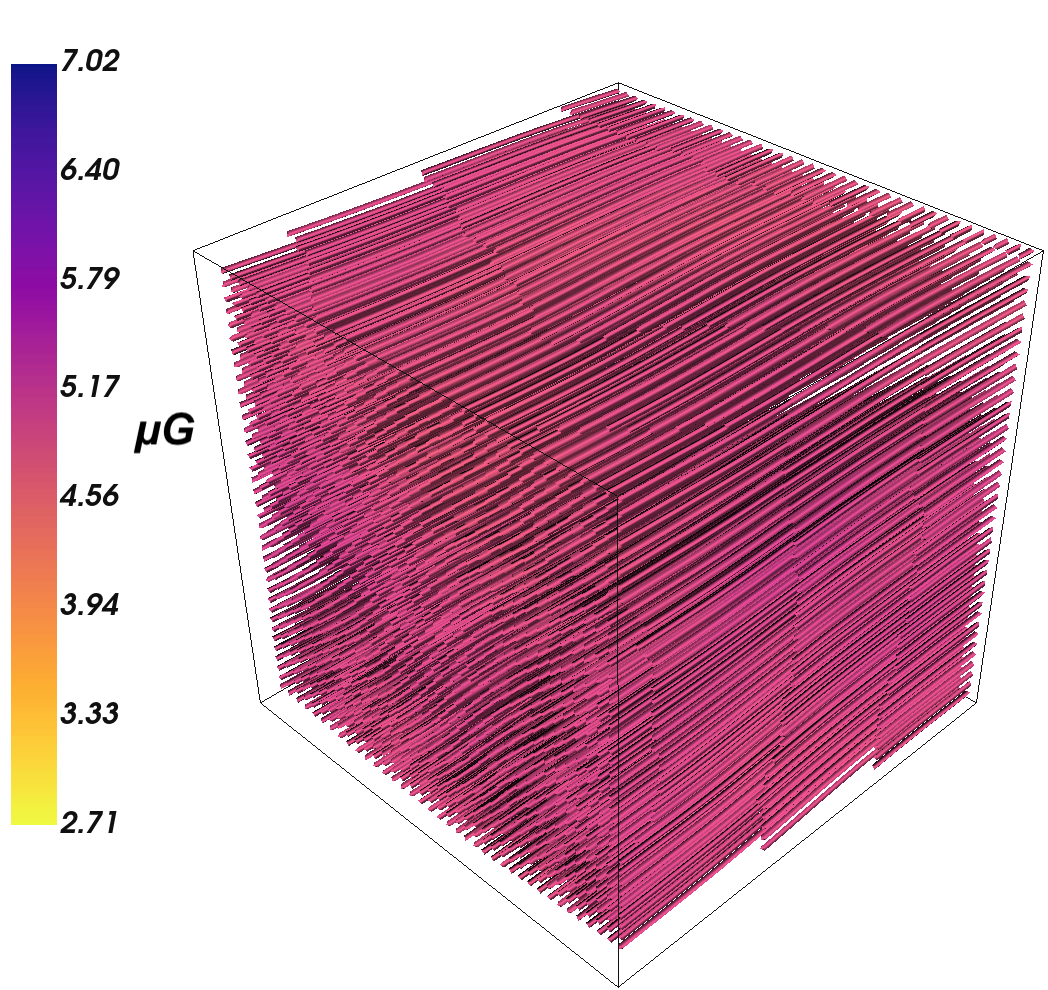}
         \caption{}
         \label{fig:10c}
     \end{subfigure}
        \caption{ Reconstruction of a 3D magnetic field within a cube of side $L = 3=$ kpc; \textbf{Top:} The ground truth; a uniform field parallel to the $x-y$ plane with magnitude $5$ $\mu$G, plus a turbulent, random, field with RMS magnitude of $1$ $\mu$G, corresponding to $\lambda  = 0.2$ in equation \eqref{lambda}. \textbf{Middle:} Local data sampled randomly, with a constant mean sampling rate $(600\text{ pc})^{-1}$. The colormap is saturated at the maximum magnitude that appears in Fig. \ref{fig:10a}. \textbf{Bottom}: The mean of the approximating posterior distribution attained via the geoVI algorithmn based on the data provided in Fig. \ref{fig:10b}.}
        \label{fig:ex1}
    \end{figure}

\begin{figure}
        \centering
        \includegraphics[scale=0.33]{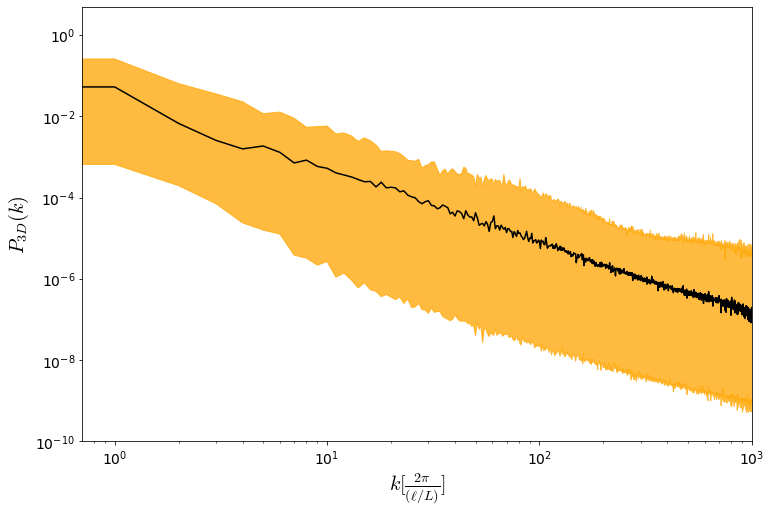}
        \caption{The three dimensional power spectrum $P_{3D}(k)$ corresponding to the example showcased in Fig. \ref{fig:ex1}. The black line is the power spectrum of the mock signal, while the orange envelope encompasses the posterior samples, providing the variance for the posterior mean power spectrum estimate. The wavevector $k$ is given in units of $2 \pi / N$. The large scatter is due to the fact that we have few and noisy datapoints, in a signal that is dominated by its uniform component, and thus the modes with $k > 0$ are poorly constrained.}
        \label{fig:pspec_ex1}
    \end{figure}

An illustrative example is shown in Fig. \ref{fig:ex1}, where we choose $k_{\text{sample}}=(600 \text{ }pc)^{-1}$ as the mean sampling rate, and a turbulent-to-uniform ratio $\lambda = 0.2$ (see equation \eqref{lambda}). In this case, the posterior mean essentially identifies the uniform component (Fig. \ref{fig:10c}) provided the sparse and local data shown in Fig. \ref{fig:10b}. The dominance of the ordered component results in a large scatter in the power spectrum (Fig. \ref{fig:pspec_ex1}). To see why this is the case, note that the zero mode of the field is 

\begin{equation}
    \hat{\B}(0) = \int \Bx d^3 \xbf \propto \B_0.
\end{equation}
In the scenario of a strong uniform component, the modes $\hat{\B}(\kbf)$ with $k > 0$ are poorly constrained by the data, which mainly inform with respect to the zero mode.


It is important to note, however, that the UHECR's deflection will primarily be influenced by the uniform component (or the zero mode), and will be affected much less by the small scale, turbulent fluctuations. Therefore, for the purposes of our inquiry, the recovery of the zero mode is enough to provide the leading order correction on the original arrival direction of UHECRs. In the next section, we will apply the geoVI algorithm as exemplified in Fig. \ref{fig:ex1} specifically to the case of backtracking UHECRs back to their sources by utilising local data.

\subsection{Correcting the arrival directions of UHECRs} \label{section:Correcting}

In the previous section we demonstrated the results of the geoVI algorithm applied to sparse and local measurements of a magnetic field for the case of a strong uniform component. However, the quality of the reconstruction should be judged in context, as different degrees of resolution are required in different applications. In fact, in this context, the GMF reconstruction is a means to an end, the latter being correcting for the effect of the GMF on the arrival direction of the UHECRs, given their arrival direction and energy as observed on Earth. \par
Stated explicitly, \emph{given an observed arrival direction $\mathbf{\hat{v}}^{\text{obs}}$ of a charged particle of known lab frame energy and charge, what is the posterior distribution for its original, extragalactic direction assuming we are provided with sparse and local measurements of the GMF within a region $\mathcal{V}$ within which it traveled?} Since the geoVI algorithm returns as output sample configurations of the GMF drawn from the posterior, we can reconstruct the UHECR's path through each sample by solving the equations of motion backwards (for details on how this is carried out in practice, see Appendix \ref{appendix:2}). In the end, we are left with a distribution for possible paths, all converging to the velocity observed on Earth. 

\begin{figure*}
    \centering
        \includegraphics[scale = 0.23]{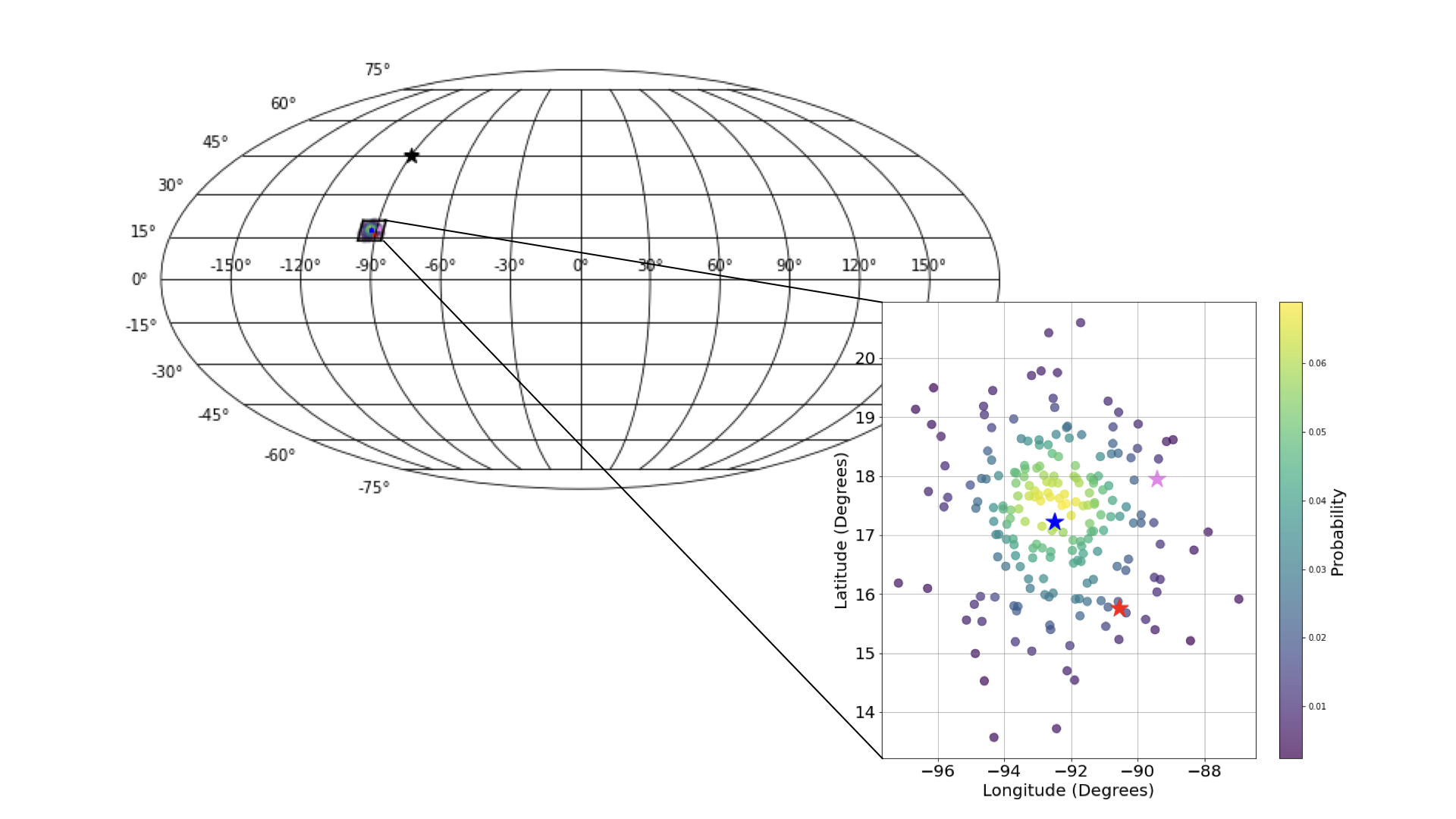}
        \caption{Plane of the sky projections of the UHECR arrival directions as inferred given local and sparse GMF data based on $100$ posterior samples (viridis colormap). The GMF reconstruction problem is the one shown in \ref{fig:ex1}. The red star denotes the true arrival direction, while the black star denotes the observed arrival direction, and they are found $1.6 \sigma$ and $22.8 \sigma$ from the posterior's mean, respectively. For comparison, we also infer the UHECR arrival direction using a simple vector mean of all the GMF data points (blue star), as well as a a nearest neighbour estimate (pink star), where at each point the GMF is assumed to be that dictated by the nearest available data point. Note that these simple reconstruction methods do not provide an uncertainty quantification. In addition, they tacitly assume either a low $\lambda$, or data points that are populated densely enough - something that cannot be expected from the distribution of dust clouds in real life applications.}
        \label{fig:Galactic_coords_Posterior_aniso}
    \end{figure*}

For the situation presented in Fig. \ref{fig:ex1}, we consider a particle of charge $e$ with final position $\mathbf{r}_i = (L/2,L/2,0)^T$ (Earth's location inside $\mathcal{V}$), velocity direction $\mathbf{\hat{v}}^{\text{obs}} = (-90^\circ, 45^\circ)$ in Galactic coordinates, and energy in the lab frame $E = 5 \times 10^{19} \text{ }eV$ is backtracked through the true GMF, as well as samples of the posterior distribution of the GMF given the local and sparse data shown in Fig. \ref{fig:10b}. In Fig. \ref{fig:Galactic_coords_Posterior_aniso}, in Galactic coordinates, the observed arrival direction (black star), true arrival direction (red star), and posterior samples (dots in viridis colormap), are shown. The colormap corresponds to the mean posterior distribution as inferred from our samples using the IFT based density estimator DENSe (\citealt{DENSe}). The advantage of this method is that the parameters of the kernel are inferred instead of assumed, and the result of this algorithm are samples of possible distributions drawn from the posterior distribution of distributions, given the samples\footnote{In addition to the reference provided, the reader is advised to look at \hyperlink{https://ift.pages.mpcdf.de/public/dense/}{https://ift.pages.mpcdf.de/public/dense/} for further information.}. From the former, the mean is drawn as our estimate for the underlying distribution, shown in the colormap of Fig. \ref{fig:Galactic_coords_Posterior_aniso}.

It can be seen that we are able to substantially correct for the effect of the GMF. For the purpose of comparison, we also include the respective result obtained via two different, much simpler, reconstruction methods: 1) merely taking the vector mean of the data points (blue star), and 2) a nearest neighbour estimate wherein one assumes that at each point in space, the value of the GMF is that given by the nearest available data point (pink star). While all three of the methods are able to correct for the effect of the GMF by essentially picking out the zero mode - which in this case predominantly affects the UHECR paths - the IFT based method is a statistically rigorous way to perform the inference, as it also provides a quantification of the inference's uncertainty, a feature that the simpler methods lack. In addition, the simple methods assume either a low $\lambda$, or data points that are populated densely enough - something that cannot be expected from the distribution of dust clouds in real life applications.

In order to quantify the improvement in our knowledge, we use the Mahalanobis distance (\citealt{mahalanobis}) between a given arrival direction $\hat{\mathbf{v}}$, and the posterior samples acquired from the geoVI algorithm,

\begin{equation} \label{Mahalanobis}
d_M[\hat{\mathbf{v}} ; P] = \sqrt{(\hat{\mathbf{v}} - \hat{\mathbf{\mu}})^{T} S^{-1}(\hat{\mathbf{v}} - \hat{\mathbf{\mu}})},
\end{equation}
where $S^{-1}$ and $\hat{\mathbf{\mu}}$ are the inverse covariance matrix and mean of $P(\hat{\mathbf{v}} | d)$, respectively. Essentially, $d_M$ measures how many standard deviations $\hat{\mathbf{v}}$ is located away from $\hat{\mathbf{\mu}}$. If  $\mathbf{\hat{v}}^{\text{true}}$ and $\mathbf{\hat{v}}^{\text{obs}}$ are the true and observed arrival directions respectively shown in Figs \ref{fig:ex1} and \ref{fig:Galactic_coords_Posterior_aniso}, we calculate $d_M[\hat{\mathbf{v}}^{\text{true}} ; P] = 1.6 \sigma$ and $d_M[\hat{\mathbf{v}}^{\text{obs}} ; P] = 22.8 \sigma$.
    
We also quantify our results using a different metric. Let $\mathbf{\hat{v}}^{\text{rec}}$ denote the arrival direction as obtained by backtracking through a reconstructed magnetic field. If we define the angle

\begin{equation} \label{angle}
    \theta' = \cos^{-1} (\mathbf{\hat{v}}^{\text{true}} \cdot \mathbf{\hat{v}}^{\text{rec}}),
\end{equation}
we are interested in the posterior mean of $\theta'$, henceforth referred to as $\theta \equiv \langle \theta' \rangle$.

\subsection{Effect of the sampling rate}

\begin{figure}
    \centering
    \begin{subfigure}{.5\textwidth}
     \centering
     \includegraphics[scale = 0.45]{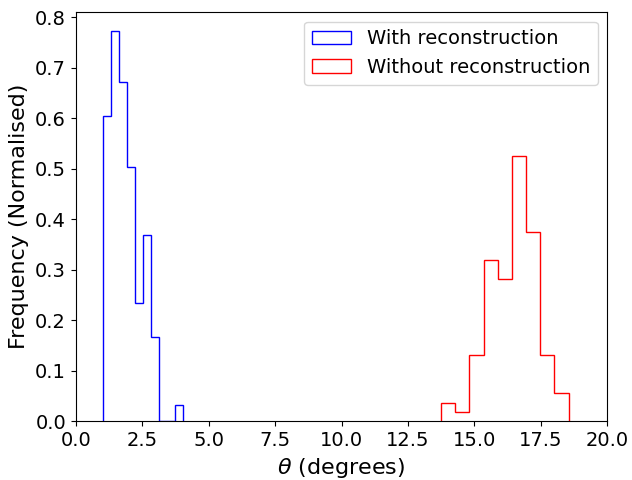}
     \caption{$k_\text{sample} = (600 \text{ }\text{pc})^{-1}$}
     \label{fig:hist_125}
 \end{subfigure}
 \hfill
 \begin{subfigure}{.5\textwidth}
     \centering
     \includegraphics[scale = 0.44]{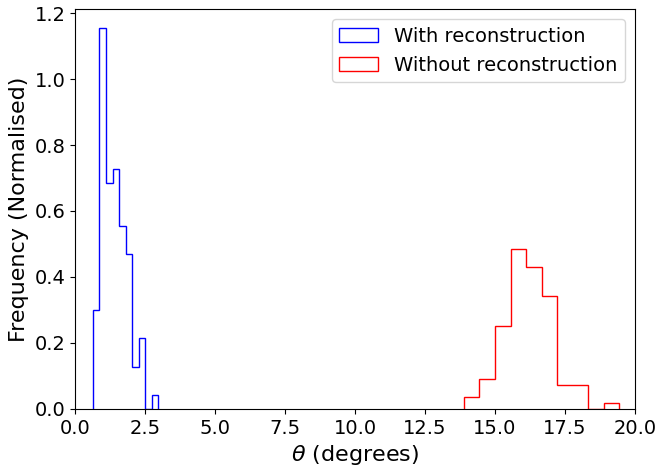}
     \caption{$k_\text{sample} = (425 \text{ }\text{pc})^{-1}$}
     \label{fig:hist_350}
 \end{subfigure}
 \hfill
 \begin{subfigure}{.5\textwidth}
     \centering
     \includegraphics[scale = 0.44]{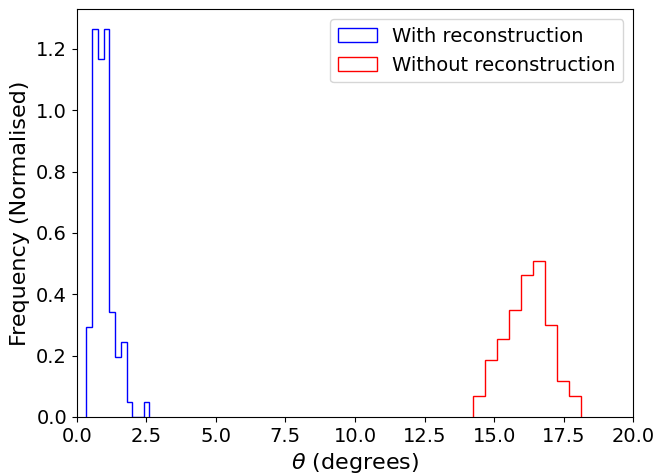}
     \caption{$k_\text{sample} = (300 \text{ }\text{pc})^{-1}$}
     \label{fig:hist_1000}
 \end{subfigure}
    \caption{Histograms of the deviation angle between the true arrival direction of UHECRs, and the one obtained by backtracking through the magnetic field as reconstructed using local data sampled at three different mean rates $k_\text{sample} = (600 \text{ pc})^{-1}$, $(425 \text{ pc})^{-1}$, and $(300 \text{ pc})^{-1}$ (blue). For each mean sampling rate, we consider $100$ independent GMFs. The observed arrival direction, the charge, and lab frame energy are assumed to be known and are $(0,0,-1)^T$, $+e$, and $5 \times 10^{19} \text{ }eV$, respectively. The turbulent-to-uniform magnitude ratio (eq. \eqref{lambda}) is set at $\lambda = 0.2$. For reference, we also plot the same angle by completely neglecting the effect of the GMF on the particle's trajectory (red). It can be seen that local and homogeneously distributed GMF data significantly improve our estimate of the true arrival directions of UHECRs, and that there is successive improvement as $k_\text{sample}$ becomes larger, as expected.}
    \label{fig:histograms1}
\end{figure}

In order to see how local sampling of the GMF might help us resolve the UHECR initial directions in general, the same backtracking procedure that led to Fig. \ref{fig:Galactic_coords_Posterior_aniso}, is carried out for an ensemble of $100$ GMF configurations, for three different values of the mean sampling rate; $k_\text{sample} = (300 \text{ pc})^{-1}$, $(425 \text{ pc})^{-1}$, and $(600 \text{ pc})^{-1}$, and each time a different field serves as ground truth and is reconstructed, via the process described in section \ref{sec:gp}. The angle $\theta$ defined in the end of the last section is calculated for each case, and the histograms for each sampling rate considered are shown in blue, in Fig. \ref{fig:histograms1}. For reference, we also calculate the angle between $\mathbf{\hat{v}}^{\text{true}}$ and the final velocity direction, $\mathbf{\hat{v}}^{\text{obs}}$, 

\begin{equation} 
    \theta_0 = \cos^{-1} (\mathbf{\hat{v}}^{\text{true}} \cdot \mathbf{\hat{v}}^{\text{obs}} ).
\end{equation}
The histogram for $\theta_0$ is shown in red for each sampling rate considered. At zeroth order, without attempting to reconstruct the GMF at all, the UHECR's path would not be altered, and so in this case $\mathbf{\hat{v}}^{\text{true}} \simeq \mathbf{\hat{v}}^{\text{obs}}$ or $\theta \simeq \theta_0$, trivially. Therefore, $\theta_0$ is the zeroth order approximation, without any data points considered. As can be clearly seen, the usage of the local and sparse data $\mathbf{d}$ significantly corrects for the effect of the GMF on our knowledge of the UHECR arrival direction. Further, as expected, increasing sampling rates tend to move the PDF for $\theta$, $P(\theta)$, more towards $\theta  = 0$. In the extreme case of a perfectly known GMF, $P(\theta)$ would be a delta function centered at zero.
    
These results are for the set value of $\lambda=0.2$. As $\lambda$ becomes larger, or as the fluctuating/turbulent component $\mathbf{b}(\xbf)$ in eq. \eqref{uni+fluct} becomes larger, our results should become increasingly worse. The reason for this is that if $\lambda \ll 1$ then the uniform component dominates, and we are required to only infer that, as the fluctuating part will only play a subdominant role in the deflection of the UHECR. As $\lambda$ becomes larger, then the process will need to discern more irregular structures that play an increasingly important role in deflecting the UHECR. This task is more difficult to achieve with sparse data. Additionally, the larger $\lambda$ becomes the less the UHECR path is expected to be deflected, as for dominating $\mathbf{b}(\xbf)$ (which has zero mean) the UHECR will travel through regions that will partly cancel each other's effect on the path resulting on a random walk, rather than a systematic deflection, about the true position of the source on the sky. Therefore, in Fig. \ref{fig:histograms1}, as $\lambda$ increases, we expect $P(\theta)$ (blue histogram) to diffuse towards larger $\theta$ (owing to the increasing resolution that is required in that case). Conversely, the PDF for the angular distance between the final velocity and the true initial velocity  (red histograms) is expected to shift towards smaller $\theta$.

\begin{figure}
    \centering
    \begin{subfigure}{.5\textwidth}
     \centering
     \includegraphics[scale = 0.45]{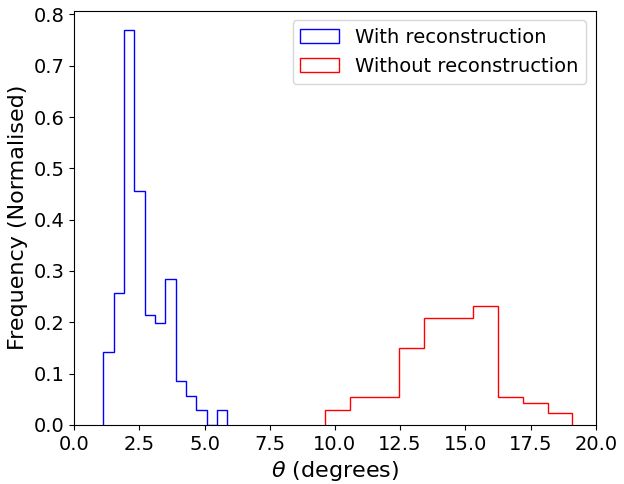}
     \caption{$\lambda = 0.5$}
     \label{fig:hist_125}
 \end{subfigure}
 \hfill
 \begin{subfigure}{.5\textwidth}
     \centering
     \includegraphics[scale = 0.43]{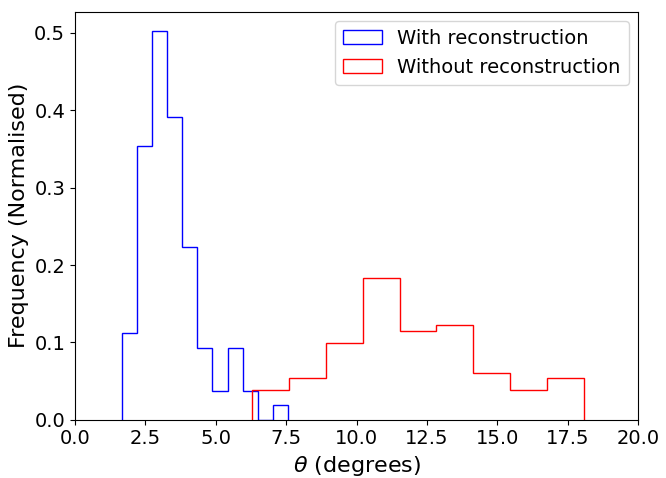}
     \caption{$\lambda = 1$}
     \label{fig:hist_350}
 \end{subfigure}
 \hfill
 \begin{subfigure}{.5\textwidth}
     \centering
     \includegraphics[scale = 0.45]{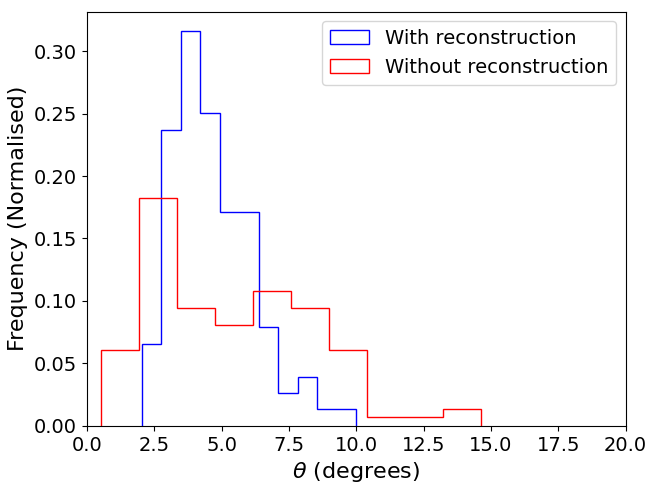}
     \caption{$\lambda = \infty$}
     \label{fig:hist_1000}
 \end{subfigure}
    \caption{Histograms of the deviation angle between the true arrival direction of UHECRs, and the one obtained by backtracking through the magnetic field as reconstructed using local data sampled at three turbulent-to-uniform ratios $\lambda = 0.5$, $\lambda = 1$, and  $\lambda = \infty$, for $\lambda$ defined in eq. \eqref{lambda}. For value of $\lambda$, we consider $100$ independent GMFs. The observed arrival direction, the charge, and lab frame energy are assumed to be known and are $(0,0,-1)^T$, $+e$, and $5 \times 10^{19} \text{ }eV$, respectively. The sampling rate (eq. \eqref{ksample}) is set at $k_\text{sample} = 600 \text{ pc}$ . For reference, we also plot the same angle by completely neglecting the effect of the GMF on the particle's trajectory (red). It can be seen that as $\lambda$ becomes larger, the deviations of the original direction as obtained by backtracking through the reconstructed GMF and the true arrival direction become more dispersed, while the observed discrepancy diffuses towards smaller angles.}
    \label{fig:histograms2}
\end{figure}

In Fig. \ref{fig:histograms2} we test this expectation, by performing the same calculation as in Fig. \ref{fig:histograms1}, but this time with increasing $\lambda$ and keeping $k_\text{sample} = (600 \text{ } \text{pc})^{-1}$ fixed. The results confirm our intuition. We notice however, that even in the limit where the turbulence is completely isotropic, $\lambda \rightarrow \infty$, the PDF resulting from considering the reconstruction is narrower than the the one where only observed directions are used. 

\begin{figure}
        \centering
        \begin{subfigure}{.5\textwidth}
         \centering
         \includegraphics[scale = 0.2]{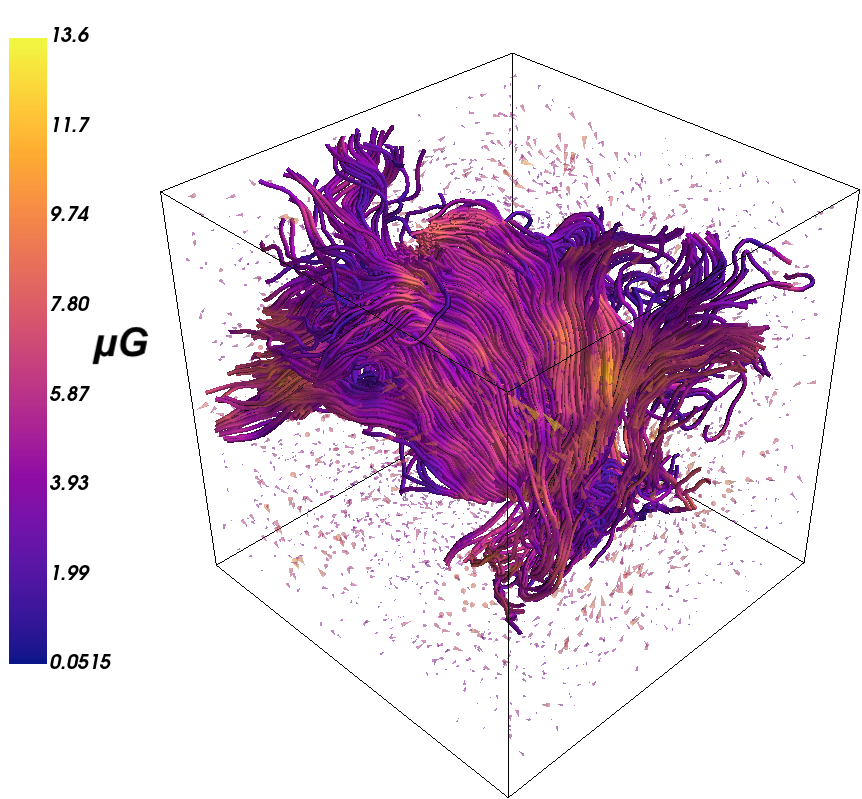}
         \caption{}
         \label{fig:ex2a}
     \end{subfigure}
     \hfill
     \begin{subfigure}{.5\textwidth}
         \centering
         \includegraphics[scale = 0.2]{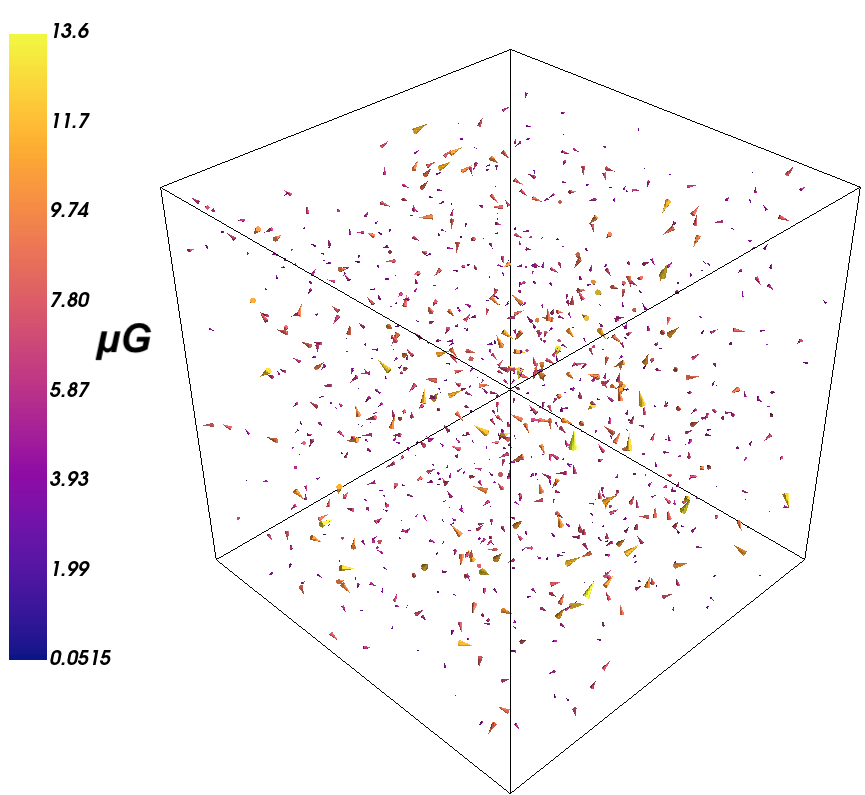}
         \caption{}
         \label{fig:ex2b}
     \end{subfigure}
     \hfill
     \begin{subfigure}{.5\textwidth}
         \centering
         \includegraphics[scale = 0.2]{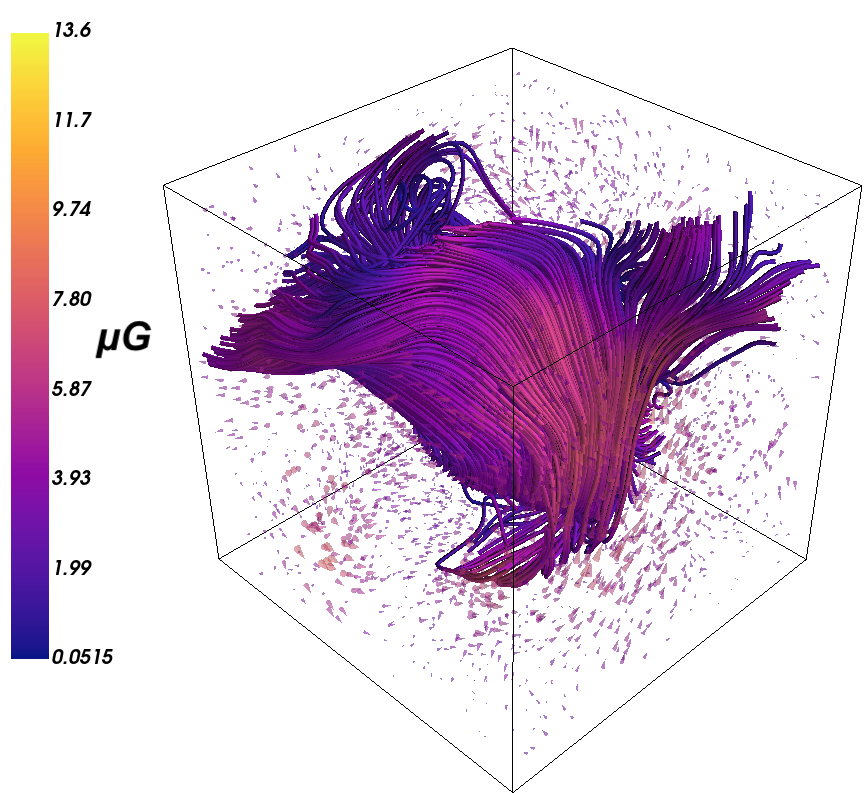}
         \caption{}
         \label{fig:ex2c}
     \end{subfigure}
        \caption{Reconstruction of a isotropically turbulent 3D magnetic field within a cube of side $L = 3 \text{ }kpc$. \textbf{Top}: the ground truth; an isotropically turbulent GMF with RMS magnitude of $5 \text{ }\mu$G. Colormap denotes the vector's magnitude. \textbf{Middle}: Local data sampled uniformly. Mean distance between each datapoint is $300 \text{ }\text{pc}$. The colormap is saturated at the maximum magnitude that appears in Fig. \ref{fig:ex2a}. \textbf{Bottom}: The mean of the approximating Gaussian posterior distribution attained via the geoVI algorithmn based on the data provided in Fig. \ref{fig:ex2b}.}
        \label{fig:ex2}
    \end{figure}

To see this, consider raising $k_\text{sample}$ to $(300 \text{ } \text{pc})^{-1}$ which is not an unrealistic expectation for the density of HI clouds relatively close to the Galactic disk. Then, for a strongly turbulent case $\lambda \rightarrow \infty$, our reconstruction can be seen in Fig. \ref{fig:ex2}. The mean of the posterior as found by the geoVI algorithm, given the data of Fig. \ref{fig:ex2b}, can be seen in Fig. \ref{fig:ex2c}. In this example, the mean is essentially a `combed' version of the true field, in that it does contain all the main structures and larger-scale features, but completely misses out the fluctuations below a certain length scale. It should be made clear, however, that the result of the reconstruction - as is the case with any inference problem - is not just the mean but rather the whole posterior distribution. The mean is drawn as a representative example of the posterior distribution, but in general samples from the whole posterior are utilised in backtracking the cosmic rays and thus obtaining the posterior for the initial arrival directions. 

\begin{figure}
    \centering
    \includegraphics[scale=0.33]{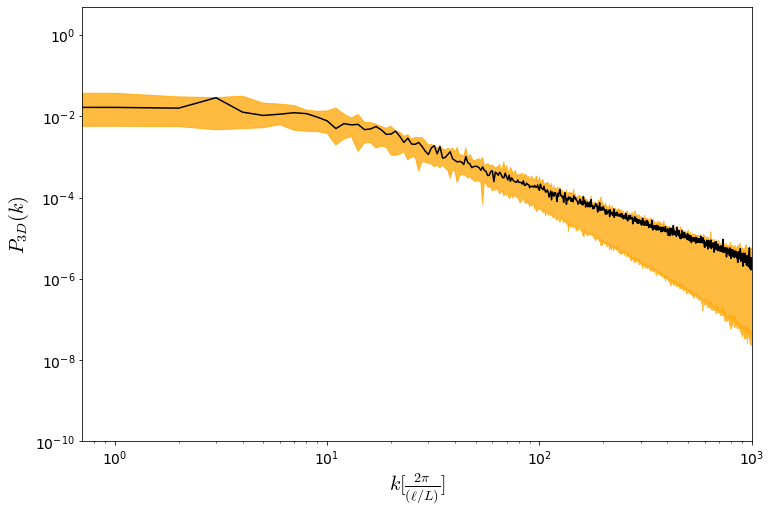}
    \caption{As with Fig. \ref{fig:pspec_ex1}, but for the case displayed in Fig. \ref{fig:ex2}. In this case, all modes are relevant and well constrained, as opposed to the case of Fig. \ref{fig:pspec_ex1}. Further, note that as $k$ becomes larger the posterior of the power spectrum starts deviating from that of the signal, implying a loss of information at smaller lengthscales, as observed in Fig. \ref{fig:ex2}.}
    \label{fig:pspec_ex2}
\end{figure}

In Fig. \ref{fig:pspec_ex2}, the power spectrum defined via equation \eqref{2pointcor}. We notice that in the case of isotropic turbulence, the modes with nonzero wavevector are constrained much better from the data compared to the anisotropic case of a dominating zero mode. 

\begin{figure*}
    \centering
        \includegraphics[scale = 0.23]{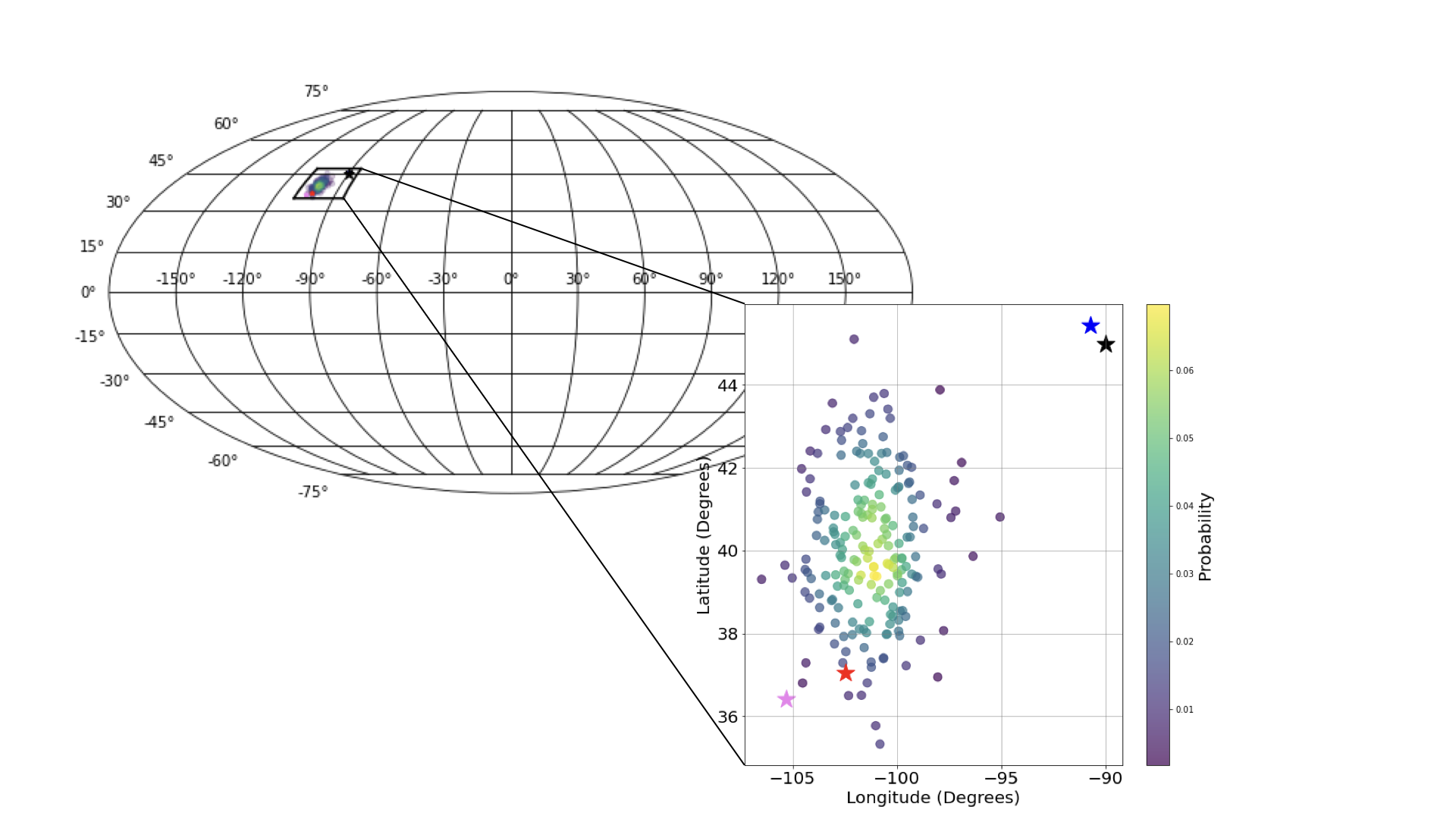}
        \caption{As Fig. \ref{fig:Galactic_coords_Posterior_aniso}, but for the turbulent magnetic field case displayed in Fig. \ref{fig:ex2}.}
        
        \label{fig:Galactic_coords_Posterior_iso}
    \end{figure*}

As was done with the weakly turbulent example, consider a UHECR with charge $e$, observed arrival direction $\mathbf{\hat{v}}^{\text{obs}} = (-90^\circ, 45^\circ)$ in Galactic coordinates, and energy in the lab frame $E = 5 \times 10^{19} \text{ }eV$ through this strongly turbulent field. In Fig. \ref{fig:Galactic_coords_Posterior_iso}, it can clearly be seen that in the extreme case of completely isotropic turbulence, there is still a substantial deviation due to the presence of the GMF, and the zeroth order approximation (that is, assuming $\hat{\mathbf{v}}^{\text{true}} \simeq \hat{\mathbf{v}}^{\text{obs}}$) is further away from the posterior's mean than the  true arrival direction. In Fig. \ref{fig:Galactic_coords_Posterior_iso} we plot the posterior samples of the UHECR arrival directions given the local GMF data shown in Fig. \ref{fig:ex2b}. The viridis colormap is the estimated posterior distribution using DENSe (see section \ref{section:Correcting}), based on the $100$ posterior samples drawn using geoVI.  As with Fig. \ref{fig:Galactic_coords_Posterior_aniso}, the red star denotes the true arrival direction, while the black star denotes the observed arrival direction. The Mahalanobis distance between these directions and the arrival directions sampled from the posterior is computed at $1.8 \sigma$ and $6.7 \sigma$ respectively. As before, two simple GMF inference schemes are also employed, for comparison: 1) a simple vector mean of all the data points (blue star) and a nearest neighbour regression (pink star). Since in this case the GMF's zero mode does not predominantly contribute in the UHECR's deflection, the simpler vector mean inference method fails completely, as it requires a dominant zero mode, or equivalently a low $\lambda$ - a prior assumption which the geoVI method does not make. The nearest neighbour inference scheme does perform well, but this also depends on the fact that in this case our data points are dense enough. As noted before, the IFT based method improves upon the simpler methods by accounting for variation in the magnetic field and we improve upon the second by providing uncertainties (and having physically sound magnetic fields), and so the influence that a possible high $\lambda$ and/or low sampling rate will have on our confidence of the suggested extragalactic origin is unknown. The IFT based method presented in the work systematically takes care of this shortcoming.

\section{Summary, Conclusions, and Outlook}\label{sec:conclusion}
\subsection{Summary}

In this work, we used $3D$ local and sparse mock observations of the GMF scattered across a cubic domain of $3$ kpc side within the Galaxy that are statistically uniformly distributed, in order to obtain the posterior distribution for the UHECR arrival directions before they entered the domain of influence of the GMF. We considered a two component field, which was comprised of a uniform and random (turbulent) part, for various relative strengths between the two components. We used techniques from information field theory and geoVI in order to construct a systematic way for calculating sample configurations from the posterior PDF. We have adapted the correlated field model of IFT to include the case divergence-free vector field such as the GMF, so that the divergence-free condition is also taken into account during the computation of the posterior samples. 

Once the above framework was established, we used it to test the quality of our reconstruction by comparing it to the ground truth, for various cases of sampling rates and turbulent-to-uniform ratios. By backtracking a UHECR of known arrival direction and rigidity (energy per unit charge) equal to $5 \times 10^{19} \text{ } V$ through sample configurations drawn from the posterior PDF for the GMF given the sparse and local mock data, we calculated their reconstructed direction, thus producing a posterior distribution for the true arrival directions. For an ensemble of GMFs sharing the same turbulence-to-uniform ratios and the same mean sampling rate, we were able to estimate the PDF of the mean angle between the true arrival direction and the ones obtained by backtracking through posterior samples, $\theta$, \emph{and we were able to show that for the case of weak turbulence very modest sampling rates are needed in order for the predictions to not deviate from the true arrival direction by more than a few degrees in relative angle}. In particular, for a mean distance of $600$ pc between each measurement, the vast majority of the relative angles between the inferred arrival directions and the true ones are below $\sim 3^ \circ$, for the cases considered. For increasing sampling rates our results become increasingly better, as should be expected. For reference, completely neglecting the information provided by the local and sparse data, thus also neglecting the effect of the GMF, results in a relative angle between the observed arrival direction and the true one ($\theta_0$) with mean value $\theta_0 \sim 15^\circ$.

As soon as the turbulent component starts dominating over the uniform one, our results deteriorate, and in the extreme case of fully isotropic turbulence there is a significant overlap between the PDF for $\theta$ and $\theta_0$. However, even in this extreme and unrealistic case, the local and sparse GMF data still provide a significant improvement in determining the true arrival directions, and by extension the location UHECR sources. 

\subsection{Conclusions}

In conclusion, our results are briefly summarised as follows:

\begin{enumerate}
    \item We have developed a systematic framework, based on information field theory, that infers the posterior distribution of the true arrival directions of UHECRs, given sparse and local data for the GMF scattered within a region of the Galaxy. 
    
    \item For weakly turbulent fields and by using uniformly sampled local data of the GMF with an average signal to noise ratio of $2$, we are able to correct for the effect of the GMF to within $\theta \sim 3^\circ$, where $\theta$ is the angle between the inferred arrival direction and the true arrival direction. The required mean distance between each measurement can be as large as $600$ pc, in this case. In applications to real data, the locations of the data points will coincide with the locations of molecular clouds, and  a mean distance of $\sim 600$ pc is a rather conservative estimate.
    
    \item For the extreme, but unrealistic, case of an isotropically turbulent GMF our results are worse compared to the case of a dominating structured field, but local data can still provide substantial improvement of our knowledge on the true arrival directions of UHECRs, and hence on the identity of their sources.

    \item Other simple inference methods, specifically a vector mean of the data points or a nearest neighbour estimate of the GMF require very weak turbulence and/or high sampling rates, and even in this case they do not quantify the uncertainty of the inference in a statistically rigorous way. These shortcomings are addressed by the IFT based method introduced in this work, as it produces samples of the posterior distribution of arrival directions instead of a single estimate, and no prior assumption for the relative strength of the turbulent component or the sampling rate is required.
\end{enumerate}

\subsection{Final Comments \& Outlook}

In this work, the data points were chosen to have a constant mean sampling rate throughout the domain of study, with the mean distance between the measurements serving as an adjustable parameter. However, in real applications this parameter is not set by us, but rather by Nature, as our measurements will be localised wherever HI clouds exist inside the domain within which we wish to reconstruct the GMF. In particular, their spacing will depend on the distance from the Galactic plane. It might be the case that the assumption of statistical homogeneity of data introduces biases into our results. In future work, we will consider the case of inferring UHECRs true arrival directions, by using data from MHD simulations of Galactic evolution in Milky Way-like galaxies, where our mock observations will not be chosen at random by a predetermined distribution, but rather from the distribution of clouds as given by the simulation data. It should also be noted that $3D$ information on each synthetic data point is a rather idealised situation; real data will mainly consist of PoS information alone, often even integrated (with some structured weighting) along the LoS. However, in the small deflection case that is relevant for our problem, the PoS component is expected to dominate the UHECR path's deflection, since the observed velocity is parallel to the LoS. This issue will be adressed in detail in future work. Finally, we like to add that for any UHECR source that could be  identified, the difference between the observed and initial travel directions of the corresponding UHECRs will then provide information on the intergalactic magnetic field.

\begin{acknowledgements}
 A.T. and V.P. acknowledge support from the Foundation of Research and Technology - Hellas Synergy Grants Program through project MagMASim, jointly implemented by the Institute of Astrophysics and the Institute of Applied and Computational Mathematics. A.T. acknowledges support by the Hellenic Foundation for Research and Innovation (H.F.R.I.) under the ``Third Call for H.F.R.I. Scholarships for PhD Candidates'' (Project 5332). V.P. acknowledges support by the Hellenic Foundation for Research and Innovation (H.F.R.I.) under the ``First Call for H.F.R.I. Research Projects to support Faculty members and Researchers and the procurement of high-cost research equipment grant'' (Project 1552 CIRCE). A.T. would like to thank Vincent Pelgrims, Raphael Skalidis, Georgia V. Panopoulou, and Konstantinos Tassis for helpful tips and stimulating discussions. G.E. acknowledges the support of the German Academic Scholarship Foundation in the form of a PhD scholarship ("Promotionsstipendium der Studienstiftung des Deutschen Volkes").
\end{acknowledgements}

\bibliographystyle{aa}
\bibliography{bibliography.bib}

\begin{appendix}\label{appendix:1}
\section{Geometric Variational Inference (geoVI)}

In this appendix we provide a brief step-by-step overview of the geoVI algorithm, which is the main algorithm used to approximate the posterior distribution of the magnetic field, given sparse and local data. The idea is to approximate the true posterior, $P$, with an approximate one, $Q$. The approximate posterior $Q$ is chosen such that the Kullback-Leibler divergence (\citealt{KLdiv}) 
\begin{equation} \label{KL}
    D_{KL}(Q,P) \equiv \int dQ \log \left( \frac{Q}{P}\right)
\end{equation}
between the actual posterior $P$ and an approximate posterior $Q$ is minimized. The main idea of geoVI is to achieve this minimization in a new coordinate system, chosen such that $P$ - in the new coordinate system - locally closely resembles a normalized standard distribution. Once this is done, the approximating posterior $Q$ is chosen to be of the form \eqref{Gaussian_vec}. Then, the mean and covariance are chosen as the parameters with respect to which the KL divergence is minimized.

\begin{enumerate}
\item First, a coordinate transformation $\bfphi=f(\bm{\xi})$ is performed, such that the new \emph{prior} is Gaussian with unit covariance and zero mean, the \emph{standardized coordinate system}. Henceforth, the measure $\mathcal{D} \bm{\xi}$ signifies integration over all possible configurations of the vector field $\bm{\xi}$. 

\item In this new coordinate system, we calculate the Fisher information metric (\citealt{amari_2016}) $\mathcal{M}(\bm{\xi})$ for the likelihood, marginalising the data and joining it with the unit prior metric, $\mathbb{1}$. Intuitively, this may be regarded as a metric over the statistical manifold associated with the likelihood $P(d|\bm{\xi})$.

\item We seek another transformation $g$ on top of the original $f$, that turns $\mathcal{M} + \mathbb{1}$ into the Euclidean metric, locally. The motivation is that in this coordinate system, since the geometry of the statistical manifold is as simple as possible, the original posterior is more likely to be accurately described by a Gaussian. The accuracy of the step depends on the choice of the expansion point for the local transformation. It can be computed locally around the mean of the approximating Gaussian in this new coordinate system. 

\item The KL-divergence \eqref{KL} is minimised in the coordinate system to which $g$ maps to with respect to the mean referred to in the previous step, using a second-order quasi-Newton method, called Newton Conjugate Gradient (NewtonCG) (\citealt{Nocedal2006}). This is achieved by drawing sample configurations and using them to compute the KL divergence, minimising it with respect to the mean.
\end{enumerate}

\section{Back-propagating the UHECRs through the GMF} \label{appendix:2}
The equations of motion for a relativistic charged particle of charge $q$ in a static magnetic field in the lab frame $\B = \Bx$ are

\begin{equation} \label{eom}
    \frac{d (\gamma m \mathbf{v})}{dt} = q \mathbf{v} \times \B, 
\end{equation}
and 

\begin{equation} \label{electric}
    \frac{d (\gamma m c^2)}{dt} = 0, 
\end{equation}
where $\gamma$ is the particle's Lorentz factor, and $\mathbf{v}$ its velocity. Equation \eqref{electric} follows from the absence of an electric field. Substituting equation \eqref{electric} into \eqref{eom}, we get

\begin{equation} \label{eom}
    \frac{d\mathbf{v}}{dt} = \frac{q c^2}{E} \mathbf{v} \times \B, 
\end{equation}
where $E$ is the particle's lab-frame energy. 

If $\delta t$ is a small time interval, then the change in the velocity during the interval $\delta t$ is 

\begin{equation} \label{discrete_velocity}
    \delta \mathbf{v} = \mathbf{v}(t) - \mathbf{v}(t - \delta t).
\end{equation}
Using equation \eqref{eom}, we may write

\begin{equation} \label{discrete_eom}
    \frac{\delta \mathbf{\hat{v}}}{\delta t} = \frac{q c^2}{E} \mathbf{\hat{v}} \times \B,
\end{equation}
where we divided both sides by $|\mathbf{v}| \simeq c$, and $\mathbf{\hat{v}} \simeq  \mathbf{v} c^{-1}$ is the velocity's direction at any given time. Substituting equation \eqref{discrete_velocity} into \eqref{discrete_eom} and solving for $\mathbf{v}(t - \delta t)$, we obtain

\begin{equation} \label{iter_vel}
    \mathbf{\hat{v}}(t - \delta t) =  \mathbf{\hat{v}}(t) - \frac{Ze c^2}{E}(\mathbf{\hat{v}}(t) \times \mathbf{B}) \delta t,
\end{equation}
where $q = Z e$, with $e$ the electron charge and $Z$ the atomic number of the UHECR. 

If we are also given the position of the UHECR at time $t$, and we wish to calculate it at time $t - \delta t$, then we may write 

\begin{equation}  \label{iter_pos}
    \mathbf{r} (t - \delta t) = \mathbf{r} (t) - \mathbf{\hat{v}}(t) c \delta t,
\end{equation}
where we once again made the assumption $|\mathbf{v}| \simeq c$ throughout the particle's path.

Therefore, if we are given the position, charge, lab-frame energy, and observed arrival direction of a UHECR, we can use equations \eqref{iter_vel} and \eqref{iter_pos} iteratively in order to solve the equations of motion numerically. We choose $\delta t$ in the iterative process such that the length $c \delta t$ is equal to the our resolution; the total domain is subdivided into voxels of side length $c \delta t$, within which the GMF is assumed constant. For this work, this amounts to setting $c \delta t = 60$ pc, while the side of the total cubic domain is $3$ kpc.

Finally, once the initial arrival direction is obtained - this happens when the coordinates of the UHECR's location exceeds the boundaries of the domain - it is translated into galactic coordinates via the equations 

\begin{align} \label{PoS_convertion}
    b &= \sin^{-1} (\hat{v}_z), \\
    \ell &= \text{sign} (\hat{v}_y) \cos^{-1} \left( \frac{\hat{v}_x}{\cos (b)} \right) \nonumber,
\end{align}
where if $\ell <0$ add $2 \pi$ to avoid negative angles.

\end{appendix}

%
%

\end{document}